\documentclass[9pt,conference]{IEEEtran}

\usepackage{waspaa25} 

\usepackage{bm} 

\usepackage{graphicx}
\usepackage{subcaption}
\usepackage{float}

\title{JSQA: Speech Quality Assessment with Perceptually-Inspired Contrastive Pretraining Based on JND Audio Pairs}

\name{Junyi Fan, Donald Williamson}
\address{The Ohio State University, USA \\
\{fan.1188, williamson.413\}@osu.edu}

\begin{document}

\maketitle

\begin{abstract}
Speech quality assessment (SQA) is often used to learn a mapping from a high-dimensional input space to a scalar that represents the mean opinion score (MOS) of the perceptual speech quality. Learning such a mapping is challenging for many reasons, but largely because MOS exhibits high levels of inherent variance due to perceptual and experimental-design differences. Many solutions have been proposed, but many approaches do not properly incorporate perceptual factors into their learning algorithms (beyond the MOS label), which could lead to unsatisfactory results. To this end, we propose JSQA, a two-stage framework that pretrains an audio encoder using perceptually-guided contrastive learning on just noticeable difference (JND) pairs, followed by fine-tuning for MOS prediction. We first generate pairs of audio data within JND levels, which are then used to pretrain an encoder to leverage perceptual quality similarity information and map it into an embedding space. The JND pairs come from clean LibriSpeech utterances that are mixed with background noise from CHiME-3, at different signal-to-noise ratios (SNRs). The encoder is later fine-tuned with audio samples from the NISQA dataset for MOS prediction. Experimental results suggest that perceptually-inspired contrastive pretraining significantly improves the model performance evaluated by various metrics when compared against the same network trained from scratch without pretraining. These findings suggest that incorporating perceptual factors into pretraining greatly contributes to the improvement in performance for SQA.
\end{abstract}

\section{Introduction}
\label{sec:intro}

Speech quality assessment (SQA) is a crucial task closely related to many research topics such as speech enhancement \cite{kumar2025rlhf}, speech generation \cite{le2023voicebox}, and automatic speech recognition (ASR) \cite{Nguyen2024ExploringPS}, to name a few. Acclaimed as the gold standard for SQA, subjective listening tests that collect quality assessment scores (mean opinion scores, MOS) from humans are commonly used due to the highly reliable results rendered from these tests \cite{fan2024perspective}. Nevertheless, subjective tests are time-consuming and costly to conduct. To overcome these disadvantages, numerous objective assessment models have been proposed over the past few decades. Early efforts were mostly spent on designing instrumental algorithms such as PESQ \cite{rix2001perceptual} and POLQA \cite{beerends2013perceptual}, where the degraded audio signal is compared against its corresponding clean reference in an intrusive way. Later, non-intrusive approaches such as ITU-T P.563 \cite{malfait2006p} were proposed to assess the speech quality without references. Thanks to the rapid growth of data-driven approaches in recent years, many advanced non-intrusive models have been proposed that generalize and align better with human perceptions \cite{fu2018quality, lo2019mosnet, leng2021mbnet, yu2021metricnet, mittag2021nisqa}. Many of these networks, however, are fully supervised and require a large amount of human-annotated quality labels, which again brings back the problems of time-consuming and costly data collection.

Given the issues mentioned above, more recent approaches investigated unsupervised pretraining that do not require explicitly labeled training. For instance, pretrained models have been fine-tuned for MOS prediction tasks \cite{cooper2022generalization, ta2024enhancing}. Despite these advances, these pretrained models were not specifically designed for evaluating the perceptual aspects of speech signals. Rather, they mainly focused on linguistic features and were commonly used for related tasks such as ASR. This mismatch leaves it somewhat unclear what contributes to the models' performance in SQA, and therefore hinders possible future improvements. More recently, work by \cite{sultana2025noise} incorporated perceptual aspects into the model design, showcasing promising improvements in model performance. This motivates further investigation of the advantages of perceptually-inspired models.

To properly address these challenges, we propose JSQA, a perceptually-pretrained contrastive learning model that effectively leverages unlabeled pairs of speech signals within the just-noticeable difference (JND) level to learn perceptually-related embeddings for SQA. The motivation comes from the just noticeable difference in human perception, which can be interpreted as the smallest quality change in a signal that listeners can reliably discern \cite{mcshefferty2015just}. To achieve this, we generated pairs of audio signals by adding the same background noise from CHiME-3 \cite{barker2015third} to the same clean utterance from LibriSpeech \cite{panayotov2015librispeech} at different SNR levels within the JND range. During the contrastive pretraining stage, these audio pairs serve as the positive pairs. This forces the encoder to output similar audio embeddings for these positive pairs perceived as the same by humans, despite their subtle differences. The encoder is perceptually guided to understand the audio features similar to how humans do. This also enables it to remain invariant to other irrelevant factors such as speakers, content, and noise types. The encoder is then fine-tuned with a small subset of speech signals from the NISQA \cite{mittag2021nisqa} dataset to predict speech quality. The hypothesis in this study is that a network contrastively pretrained with perceptual JND pairs can yield embeddings that better align with human perception and can therefore reliably achieve high accuracy in MOS prediction even fine-tuned with a small set of labeled data.

\begin{figure*}[t]
    \centering
    \begin{subfigure}[b]{0.48\textwidth}
        \centering
        \includegraphics[width=\linewidth]{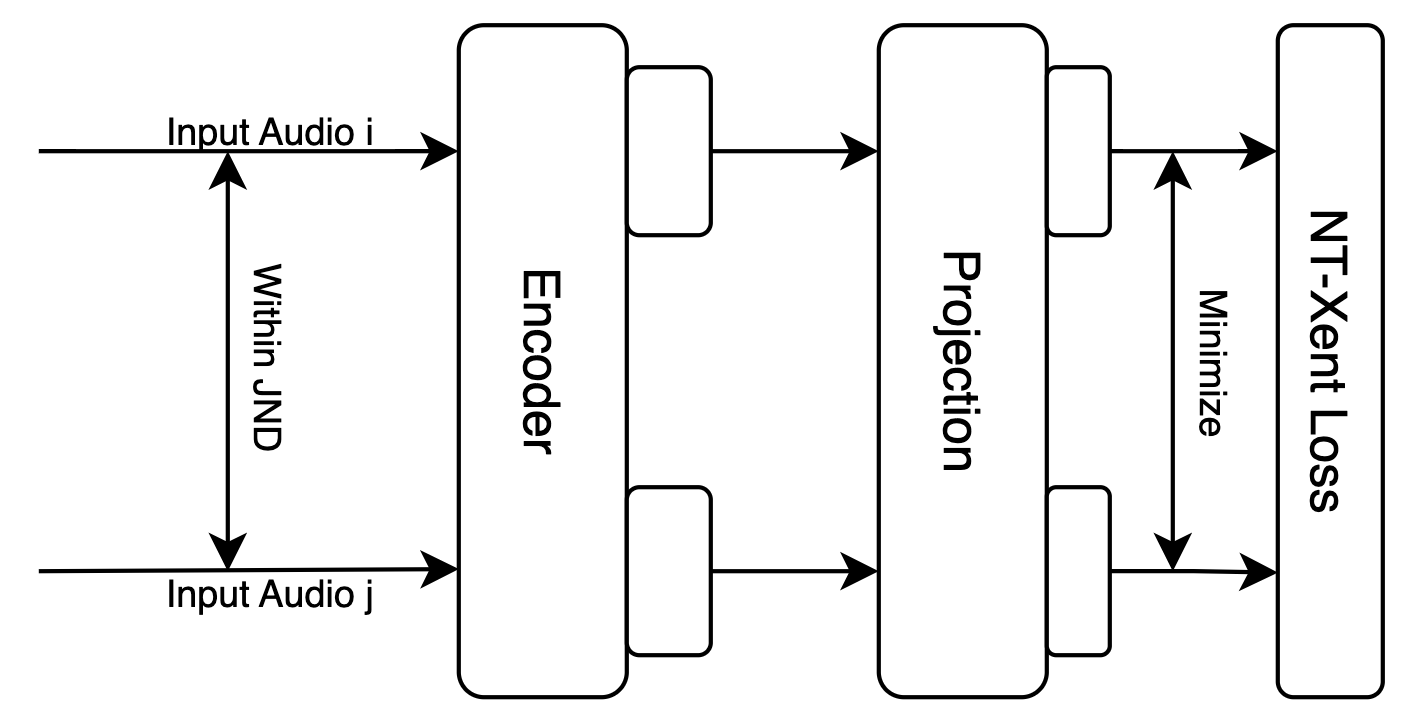}
        \caption{Contrastive pretraining for encoder. The projection head is optional.}
        \label{fig:plot1}
    \end{subfigure}
    \hfill
    \begin{subfigure}[b]{0.48\textwidth}
        \centering
        \includegraphics[width=\linewidth]{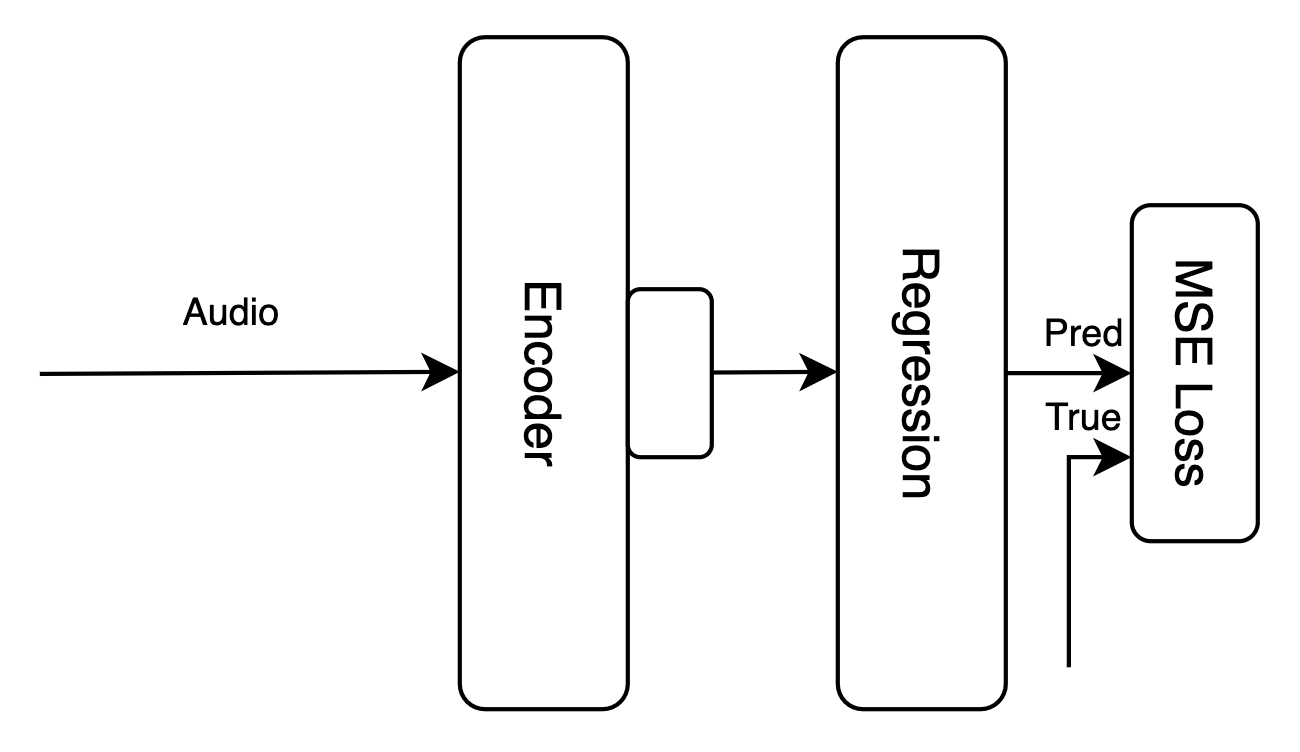}
        \caption{MOS prediction fine-tuning.}
        \label{fig:plot2}
    \end{subfigure}
    \vspace{0mm}
    \caption{Model architecture of the two-stage training. The empty boxes represent the embedding outputs from their corresponding sub-networks.}
    \label{fig:your-overall-figure}
\end{figure*}

We discuss related work in \cref{sec:Rw} and introduce the methodology in \cref{sec:M}. Experimental results are presented in \cref{sec:E}. In the end, we summarize the paper in \cref{C}.

\section{Related Work}
\label{sec:Rw}

\subsection{Traditional SQA Approaches}

Many approaches predicted quality scores with the corresponding clean reference signals. PESQ \cite{rix2001perceptual} and VISQOL \cite{visqol}, for instance, compared the degraded noisy audio signals with their references by using hand-crafted auditory features designed with signal processing techniques. These models do not correlate properly with human perception, with drawbacks such as being sensitive to minor auditory effects \cite{5133799, 6638348, asupaq}. Meanwhile, it is impractical to assume clean references are always available in many real-world situations. To overcome these issues, interest grew in studying non-intrusive methods to predict quality scores without references like ITU-T P.563 \cite{malfait2006p}. However, its use case is limited as it was specifically designed for certain systems and noise types, which worsens its performance beyond these situations.

\subsection{Data-Driven SQA Approaches}
Thanks to the development of deep learning models, groups of researchers have made efforts to utilize them for non-intrusive quality assessment. For example, Quality-Net \cite{fu2018quality} used BLSTMs to predict PESQ scores without references, with an approximate Pearson correlation coefficient of 0.9. MOS-Net \cite{lo2019mosnet} uses a CNN-BLSTM architecture to predict MOS at the system level and achieves reasonable results. Many subsequent studies such as MBNet \cite{leng2021mbnet} and MetricNet \cite{yu2021metricnet} proposed various improvements to achieve better correlations with MOS. As self-attention mechanisms grew in popularity, models such as NISQA \cite{mittag2021nisqa} have started incorporating them into quality prediction to not only further improve the performance and generalizability but also provide quality information from various dimensions, including noisiness and coloration.  More recently,  pretrained auditory models (wav2vec 2.0 \cite{baevski2020wav2vec}, HuBERT \cite{hsu2021hubert}) were fine-tuned for MOS prediction achieving superior performance \cite{Nguyen2024ExploringPS, 10626267, cooper2022generalization}, since they may provide richer embeddings learned from a large amount of unlabeled data. 

\subsection{Perceptual Training Strategy}
The current work is motivated by the idea of incorporating human perception into deep learning. DPAM, introduced by Manocha \emph{et al.} \cite{manocha2020differentiable}, is a deep perceptual audio metric learned from human-labeled JND pairs. The metric aligns well with human perceptions, which hints at the possibility of further exploiting the idea of utilizing JND pairs for various tasks to improve performance. The later CDPAM \cite{manocha2021cdpam} incorporates contrastive learning, a training strategy that has been showing promising performances in SQA \cite{nomad, ragano2024scoreq}, with augmented data and multi-dimensional representations to improve the overall performances like generalization of DPAM. The two studies came close to but did not examine the idea of leveraging the JND pairs directly in the unsupervised pretraining stage. This allows us to investigate this possibility to see whether it will bring further improvement to SQA. SESQA \cite{serra2021sesqa} utilized multi-task learning guided by a mix of real and synthetic data, part of which is a portion of the JND data from DPAM, and outperformed a fully supervised baseline by a 36\% reduction in error rate. This further affirms the assumption that perceptually guided training can benefit model performance.

\section{Method}
\label{sec:M}

In this section, we introduce the network architecture and describe the training stages.

\subsection{Model Architecture}
\subsubsection{Contrastive pretraining}

The contrastive pretraining network consists of an encoder and optionally a projection head appended after the encoder, as seen in Fig.~\ref{fig:plot1}. They both follow similar designs to the modules used in CDPAM \cite{manocha2021cdpam}, with necessary changes made to ensure they are appropriate for the current study. Specifically, the encoder is made of 16 layers of 1-D CNN of kernel size 15, batch normalization, and Leaky ReLU activation. After the last layer of convolution, a global average pooling is performed, yielding a 512-dimensional embedding. Only the first half of the embedding is used for the contrastive loss calculation to achieve better training efficiency. 

The two-layer projection head reduces the embedding dimension by half after each layer to map the embedding into a more compact latent space for contrastive loss calculation. Leaky ReLU activation and dropout are applied for the first layer. However, given that the embedding yielded by the encoder is already half the size of the one used in the original study, the projection head in this current study becomes less necessary and therefore is optionally included in the pretraining stage to investigate how including it or not may impact the model performance.

\begin{figure}[t]
    \centering
    \includegraphics[width=\linewidth]{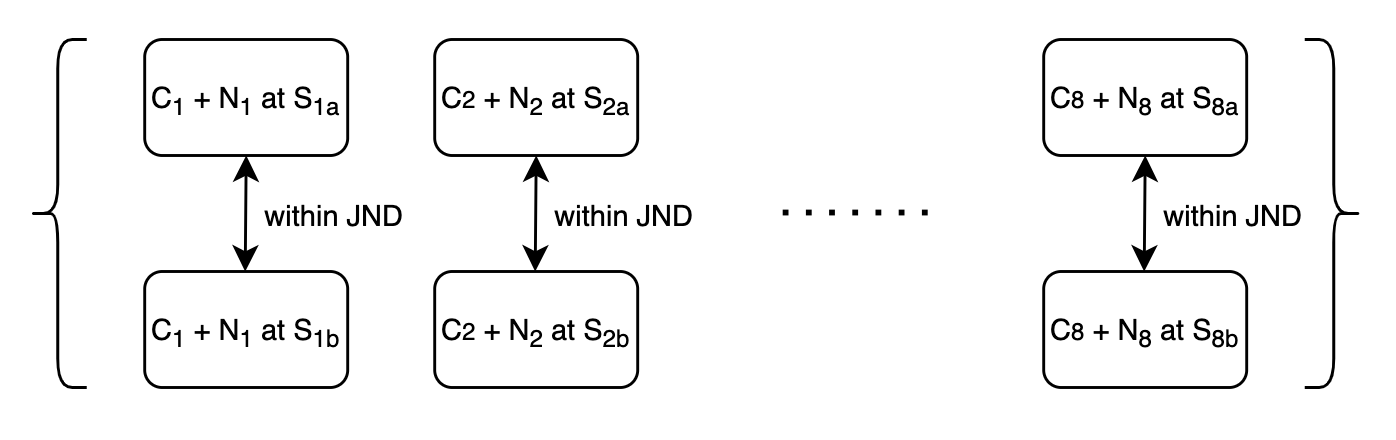}
    \caption{Illustration of JND-based audio pair construction for contrastive pretraining with a batch size of 8. $C_{\#}$ and $N_{\#}$ represent clean speech and noise signals. $S_{\#a}$ and $S_{\#b}$ denote the randomly determined SNR values at which the two generated audio signals are added.}
    \label{fig:jnd-pairs}
\end{figure}

\begin{figure}[t]
    \centering
    \includegraphics[width=0.9\linewidth]{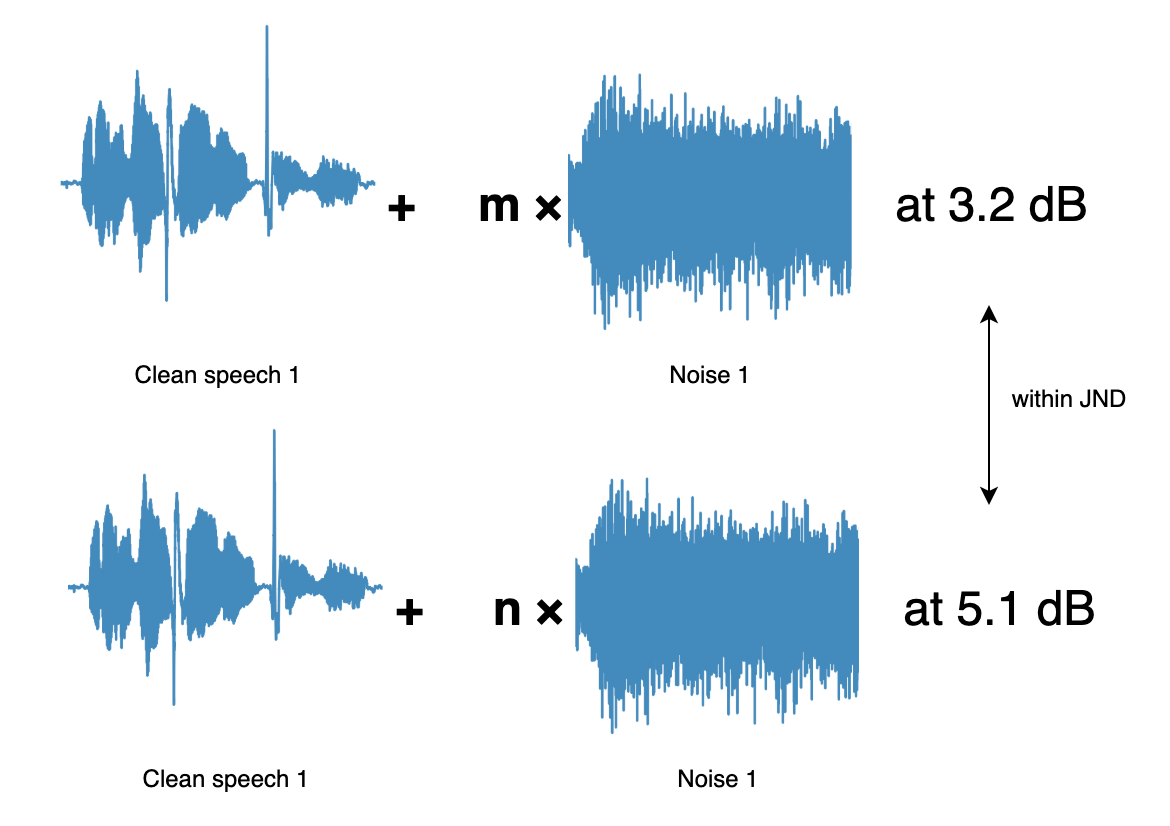}
    \caption{Illustration of a single JND pair. Specifically, $m$ is a scaling factor applied to noise 1 to ensure the SNR of the combined signal is 3.2 dB, and $n$ ensures the other combined signal is 5.1 dB. The difference of 3.2 and 5.1 dB is within the JND level.}
    \label{fig:jnd-single}
\end{figure}

The batch size for pretraining is 8 audio signal pairs, which includes 16 (8$\times$2) signals in total. Each pair, made of 2 audio signals within the JND level (see Fig.~\ref{fig:jnd-pairs} and Fig.~\ref{fig:jnd-single}), is considered as a similar pair in contrastive pretraining, where the distances of their embeddings in the latent space are minimized. Distances between any 2 audio signals from different pairs are maximized as they are treated as dissimilar pairs. The contrastive loss is borrowed from NT-Xent loss (Normalized Temperature-Scaled Cross-Entropy Loss) \cite{chen2020simple}. The loss for a positive pair $(i, j)$ is defined as:
\begin{equation}
\ell_{i,j} = -\log \frac{
    \exp\left( \mathrm{sim}(z_i, z_j) / \tau \right)
}{
    \sum\limits_{\substack{k=1 \\ k \ne i}}^{2N} \exp\left( \mathrm{sim}(z_i, z_k) / \tau \right)
},
\end{equation}
where $z_i$ is the embedding of audio $i$ in a batch of size $N$ containing $2N$ audio signals. $z_j$ is the embedding of audio $i$'s counterpart, audio $j$, which is within the JND level of audio $i$. $z_k$ is the embedding of any audio $k$ that is not audio $i$ in the batch. $\mathrm{sim}(z_i, z_j) = \frac{z_i^\top z_j}{\|z_i\| \|z_j\|}$ is the cosine similarity, and $\tau$ is the temperature parameter, where $\tau = 1$ for this study due to the small batch size.

The pretraining dataset is around 33 GB with audio sampled at 16 kHz, which is roughly only 0.5\% of the scale of full wav2vec 2.0 and HuBERT pretraining datasets. Adam is used with a learning rate of $10^{-3}$. The model was trained for approximately 45 epochs, whether the projection head was included or not, to achieve a fair comparison.

\subsubsection{MOS fine-tuning}

In the MOS fine-tuning, the pretrained encoder is connected with a regression network with MSE loss to further fine-tune the model parameters. It consists of a lightweight feed-forward network that maps the encoder embedding to a bounded range for MOS. It contains 4 fully-connected layers, each with leaky ReLU activations. In the final linear layer, the feature is reduced to a single scalar output to approximate MOS. A scaled sigmoid activation is applied after the final layer to force the range of the output score into [1, 5], the same range for MOS, which guarantees reasonable outputs. The fine-tuning network was trained for approximately 250 epochs with a batch size of 8 audio samples, with Adam used at a learning rate of $10^{-3}$.

\subsection{Pretraining and Fine-tuning with Data Design}

\subsubsection{JND estimation with SVM classifier}
\label{svm}

A previous study suggests a general order of 3 dB in SNR differences as the JND threshold in terms of speech quality \cite{mcshefferty2015just}. It conducted a series of subjective listening tests consisting of speech added with simulated noise to collect perceptual data, based on which they reached the aforementioned conclusion. As both the clean speech and noise signals in this study come from real-world recordings, this conclusion may not be suitable for producing our JND pairs used for pretraining.

To this end, we trained a simple SVM classifier that can readily predict whether a pair is within JND, based on the linearly added JND data from \cite{manocha2021cdpam}, which was effectively generated in the same way as those used in this study by adding noise to speech signals at controlled ratios. The audio pairs include JND labels that were obtained from a listening study. Therefore, the data from \cite{manocha2021cdpam} are more similar to the data used in this current study than those from \cite{mcshefferty2015just}. The SVM predicts whether a pair is within JND or not based on well-established input features like the PESQ and SI-SDR values of the signal pair, where the less noisy one of the pair is treated as the reference during PESQ or SI-SDR calculation. These features are simple and deterministic, therefore reducing the variances in the JND predictions made by the SVM, but still allowing for efficient performance. PESQ performed the best overall and was chosen as the final input feature for the classifier. 

After the SVM classifier was trained, we generated data for this study by controlling the SNR values of the signals, as it is straightforward and easy to control compared to PESQ or SI-SDR. Specifically, we randomly chose the SNR values of a pair of audio samples following a uniform distribution with a width of 6. This can be, for instance, a uniform distribution from 3 to 9. The resulting distribution of the differences of the SNR values of all pairs used for training looks like the shape shown in Fig~\ref{fig:sd}. We then used the trained SVM to validate (i.e., double-check, examine) how many pairs of the generated data are indeed within JND, judged by the SVM prediction results. Prediction results suggest 96.73\% of the final generated data are within JND, ensuring they are appropriate for pretraining.

\subsubsection{Contrastive pretraining}
\label{Cp}

\begin{figure}[t]
    \centering
    \includegraphics[width=\linewidth]{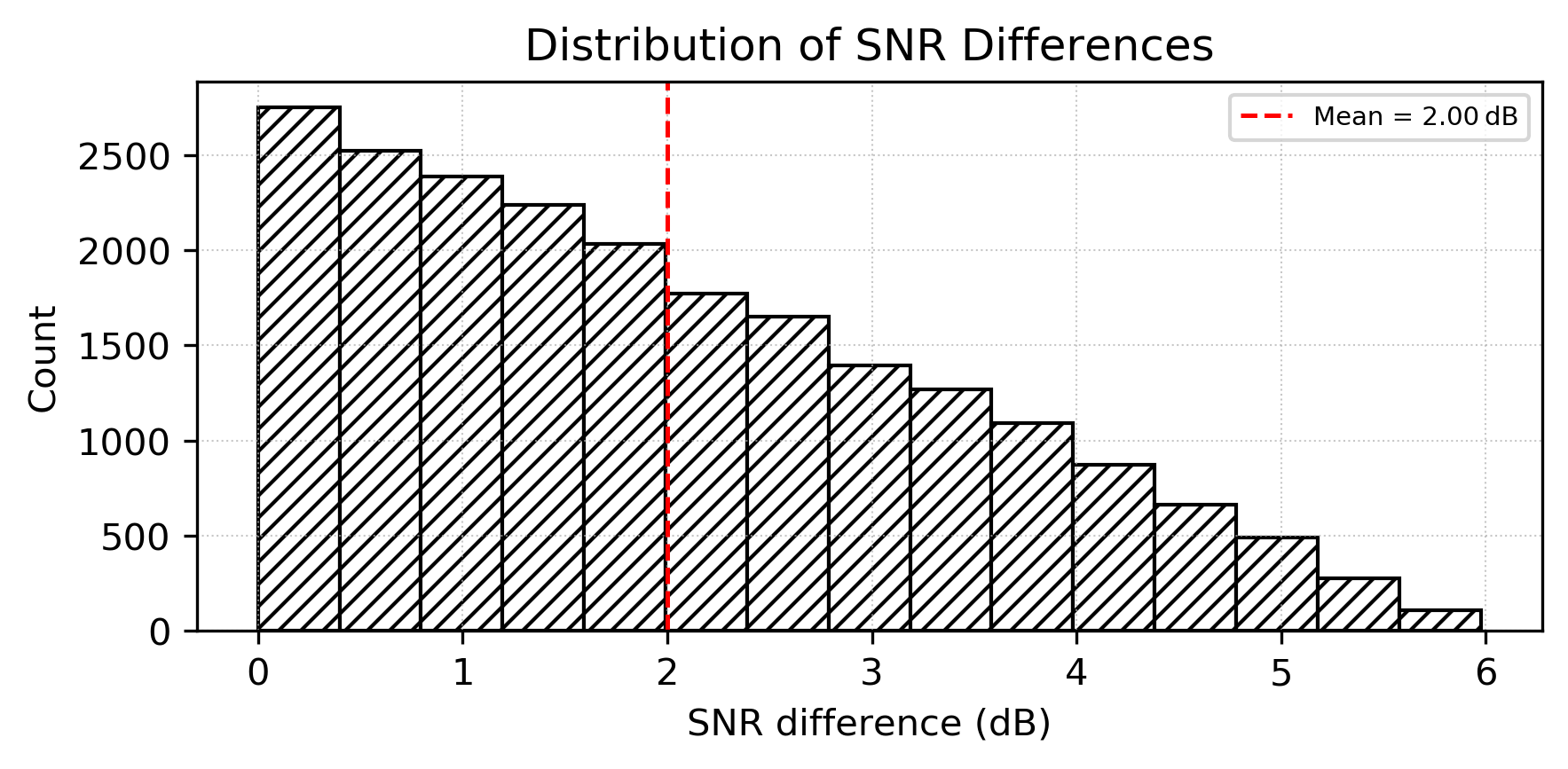}
    \caption{Distribution of the absolute difference values of the SNR levels between the JND pairs. The mean value of these differences is around 2.0 dB.}
    \label{fig:sd}
\end{figure}

To prepare the dataset for contrastive pretraining, a JND audio pair was generated by adding the same background noise from CHiME-3 to the same clean speech signal from LibriSpeech at a random SNR value within the JND level (see Fig.~\ref{fig:jnd-pairs}).
We required the SNR differences in dB of the JND pairs to be a random number between 0 and 6. This decision came from the previous study \cite{mcshefferty2015just} and the preliminary experimental results based on the classifier mentioned above in section \cref{svm}. Statistics of the generated dataset are shown in Fig.~\ref{fig:sd}, where it showcases the distribution of the absolute difference values of the SNR levels between the JND pairs. The mean value of these differences is around 2.0 dB. It should be noted that conclusions of JND thresholds made based on the feedback collected from listening study may still vary greatly. The inconsistency might be because the accurate JND thresholds vary greatly in different cases as they pertain to factors such as the listening environments, noise types, and even attention levels and listening efforts. 

The pretraining dataset contains around 33 GB audio files sampled at 16 kHz. Fig.~\ref{fig:jnd-pairs} showcases the data generation and arrangement in a batch of size 8 pairs. Fig.~\ref{fig:jnd-single} explains how each pair is generated with a concrete example, where for the same clean speech and noise sample, they are added together at different SNR values by scaling the noise samples with two scaling factors $m$ and $n$.

\subsubsection{Fine-tuning}

The small fine-tuning dataset comes from the simulated condition in NISQA dataset \cite{mittag2021nisqa}, roughly around 25 GB. The audio samples in the data are provided with their MOS evaluated by human listeners during a listening study. These data were used to fine-tune the pretrained encoder from the previous step for MOS prediction.

\section{EXPERIMENTS}
\label{sec:E}

\subsection{Encoder Training}

The encoder was trained on the data mentioned in \cref{Cp}. The number of the 1-D CNN layers is determined to be 16. The filters have a kernel size of 15, with 16 of them being used in the beginning. The number of filters increases every 4 layers. The SNR of the audio signals covers a wide range from -3 to 9 dB. PCC and SRCC are used to evaluate the correlation between the predicted scores and ground-truth scores. Higher values usually suggest better performance. RMSE and MAE are used to evaluate how much the predicted scores deviate from the ground-truth. Lower values indicate better performance.

\begin{figure}[t]
    \centering
    \includegraphics[width=0.9\linewidth]{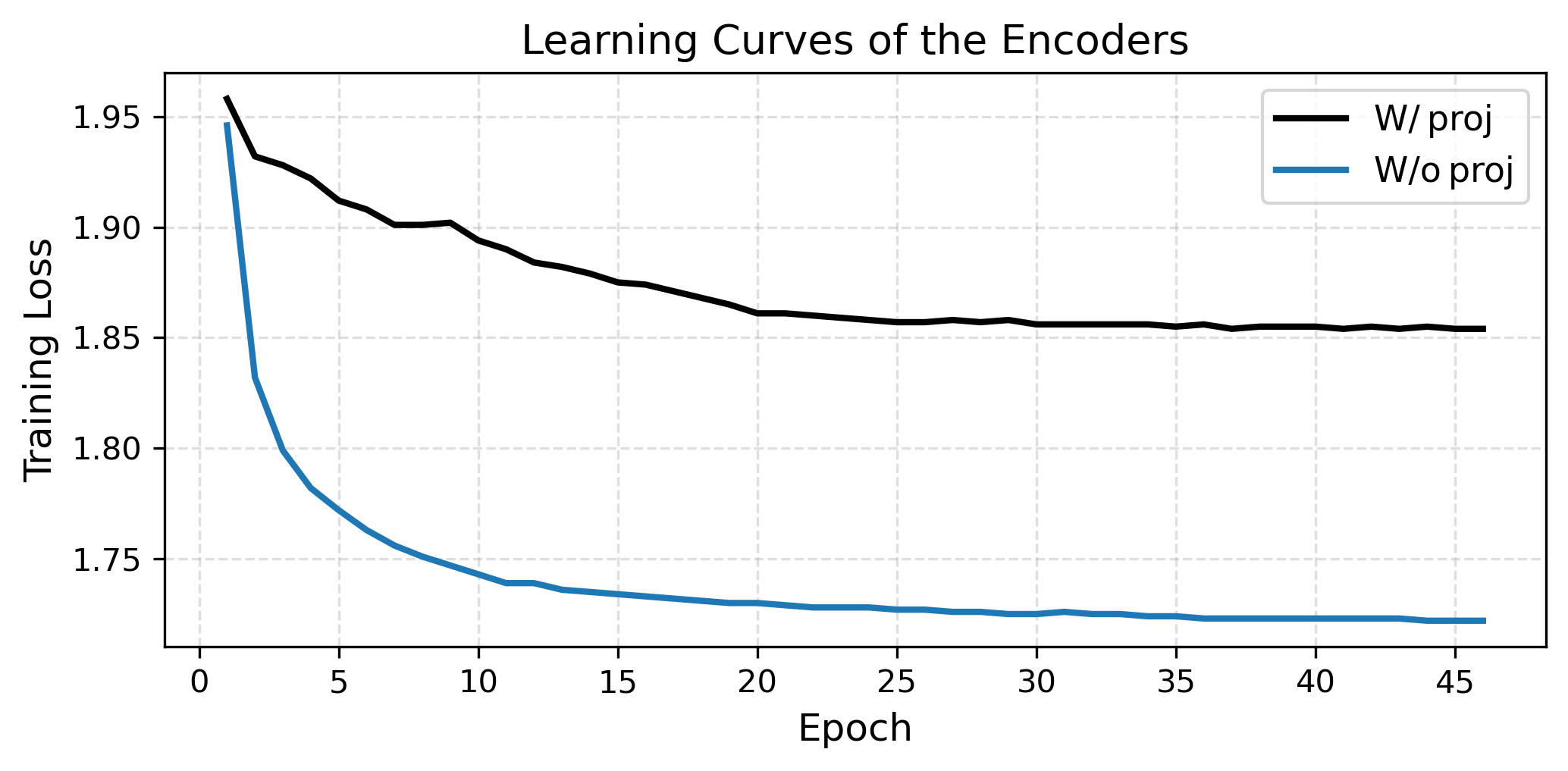}
    \caption{Learning curves for the encoder during pretraining. The upper black one and lower blue one represent the curves for the encoder with and without a projection head, respectively.}
    \label{fig:lc}
\end{figure}

Two pretraining strategies are investigated, with the only difference being whether the projection head is included. For both cases, the pretraining loss is observed to be steadily decreasing before becoming stable (see Fig.~\ref{fig:lc}). However, although trained in the same way, it can be seen that the encoder without the projection head performs better than the one with the projection head at any given epoch during the training stage based on the value of the training losses. The reason for this could be that the output embedding generated by the encoder is already appropriate for the fine-tuning stage, given its ability to well represent the learned audio features despite its compact dimensionality. Further processing the embedding and mapping it into an even more compact space through the projection head might not be necessary, despite being a more common training strategy in contrastive learning, and could even adversely affect the model performance in this specific case due to the loss of necessary information of feature representations.

It is possible that a smaller pretraining loss does not necessarily warrant better fine-tuning performance. With this in mind, both encoders pretrained with or without the projection head were further used for MOS fine-tuning to fully understand their discrepancy. The encoders both came from one of the first few epochs where the pretraining loss stopped decreasing, to ensure proper representation while avoiding overfitting.

\subsection{MOS Fine-tuning}
\label{sec:copyright}

A simple regression head was appended to the pretrained encoders for MOS prediction. Due to the advantageous pretraining that incorporates human perceptual similarity judgments into model training, the MOS prediction still performs well based on the simple network design. \cref{tab:proj-metrics-transposed} presents different metric scores to evaluate the models' performance in detail. Both encoders pretrained with or without the projection head were fine-tuned in the same way. As can be clearly seen, the result from the encoder pretrained without the projection head still outperforms the other one in every situation. This further reinforces our assumption that the projection head might not be required for this case.

\begin{table}[t]
\centering
\caption{MOS prediction metrics with and without projection head after fine-tuning. Better values per column are highlighted in bold. }
\label{tab:proj-metrics-transposed}
\begin{tabular}{lcccc}
\toprule
\textbf{Model} & \textbf{PCC} & \textbf{SRCC} & \textbf{RMSE} & \textbf{MAE} \\
\midrule

W/ proj     &     0.710    &   0.703      &    0.870     &      0.654   \\
W/o proj    &     \textbf{0.821}    &   \textbf{0.822}      &     \textbf{0.651}    &   \textbf{0.498}      \\
\bottomrule
\end{tabular}
\end{table}


Furthermore, to more thoroughly understand how beneficial the pretraining is in this study, we also conducted an experiment where the encoder was not pretrained at all while everything else remained the same. The projection head was not included after the encoder during the training. As shown in \cref{tab:ablation-metrics-transposed}, the results from the model with pretraining far surpass those trained from scratch according to all metrics. Even with a very small dataset, the pretraining significantly improves the model's performance across all metrics.

\begin{table}[t]
\centering
\caption{MOS prediction metrics with and without pretraining. Better values per column are highlighted in bold. *~lists reported results that used similar NISQA data. Italicized numbers indicate approximate values as precise figures are unavailable.}
\label{tab:ablation-metrics-transposed}
\begin{tabular}{lcccc}
\toprule
\textbf{Model} & \textbf{PCC} & \textbf{SRCC} & \textbf{RMSE} & \textbf{MAE} \\

\midrule

w2v NISQA*      &     0.780    &   0.770     &    0.585     &     -    \\

w2v 10K w/ NTsP501*      &     \textit{\textbf{0.805}}    &   \textit{\textbf{0.840}}     &    \textit{\textbf{0.446}}     &     -    \\
\midrule
W/o pretraining     &     0.712    &   0.704      &    0.795     &      0.617   \\
W/ pretraining    &     \textbf{0.821}    &   \textbf{0.822}      &     \textbf{0.651}    &   \textbf{0.498}      \\
\bottomrule
\end{tabular}
\end{table}

It is worth mentioning that the overall scale of the network also remains relatively small. Specifically, the base versions for wav2vec 2.0 have around 95M parameters, whereas our model has approximately 26M parameters, a merely 27\% of wav2vec 2.0. Its large version even has around 317M parameters, more than 12 times larger than ours. However, the performance is comparable to the model w2v NISQA from \cite{acodl} and w2v 10K w/ NTsP501 from \cite{mrh}, which is based on wav2vec 2.0 and trained on similar NISQA data. 

\section{Conclusion and Future Work}
\label{C}

This paper introduces JSQA, a perceptually-inspired two-stage training framework for non-intrusive SQA. Specifically, the first stage pretrained an encoder with contrastive loss based on JND pairs generated from LibriSpeech clean utterances and CHiME-3 background noise. The second stage fine-tuned the pretrained encoder for MOS prediction with a small set of MOS data. The key findings are: (1) perceptually guided pretraining is beneficial. JSQA decreased the prediction error by 18\% in RMSE and 19\% in MAE while increasing the PCC by 15\% and SRCC by 17\%, when compared to the same network without pretraining. (2) Small models can perform well when trained properly. Even with the improved performance, JSQA, however, still remains compact with only 26M trainable parameters and 33GB of pretraining audio data. (3) A projection head may not always be needed. One experiment suggests a possible situation where including the projection head might adversely affect model performance. In this case, when the encoder embedding is already small and well-conditioned enough, it can be directly used for fine-tuning. In the future, we may further extend the scale of the experiments, like the size of the training data to investigate how it compares with other models on similar scales. We will also look into the generalization ability of the current model, consider additional noise types, and investigate whether the current framework works well beyond additive noise.

\section{Acknowledgment}
This work was supported in part by the National Science Foundation under award number IIS-2235228.


\clearpage
\bibliographystyle{IEEEtran}
\bibliography{refs25}

\begin{thebibliography}{10}
\providecommand{\url}[1]{#1}
\csname url@samestyle\endcsname
\providecommand{\newblock}{\relax}
\providecommand{\bibinfo}[2]{#2}
\providecommand{\BIBentrySTDinterwordspacing}{\spaceskip=0pt\relax}
\providecommand{\BIBentryALTinterwordstretchfactor}{4}
\providecommand{\BIBentryALTinterwordspacing}{\spaceskip=\fontdimen2\font plus
\BIBentryALTinterwordstretchfactor\fontdimen3\font minus \fontdimen4\font\relax}
\providecommand{\BIBforeignlanguage}[2]{{%
\expandafter\ifx\csname l@#1\endcsname\relax
\typeout{** WARNING: IEEEtran.bst: No hyphenation pattern has been}%
\typeout{** loaded for the language `#1'. Using the pattern for}%
\typeout{** the default language instead.}%
\else
\language=\csname l@#1\endcsname
\fi
#2}}
\providecommand{\BIBdecl}{\relax}
\BIBdecl

\bibitem{kumar2025rlhf}
A.~Kumar, A.~Perrault, and D.~S. Williamson, ``Using rlhf to align speech enhancement approaches to mean-opinion quality scores,'' in \emph{Proceedings of the IEEE International Conference on Acoustics, Speech and Signal Processing (ICASSP)}, Hyderabad, India, 2025.

\bibitem{le2023voicebox}
M.~Le, A.~Vyas, B.~Shi, B.~Karrer, L.~Sari, R.~Moritz, M.~Williamson, V.~Manohar, Y.~Adi, J.~Mahadeokar, and W.-N. Hsu, ``Voicebox: Text-guided multilingual universal speech generation at scale,'' in \emph{Advances in Neural Information Processing Systems 36 (NeurIPS 2023)}, 2023.

\bibitem{Nguyen2024ExploringPS}
T.~Nguyen, C.~Fredouille, A.~Ghio, M.~Balaguer, and V.~Woisard, ``Exploring pathological speech quality assessment with asr-powered wav2vec2 in data-scarce context,'' in \emph{International Conference on Language Resources and Evaluation}, 2024.

\bibitem{fan2024perspective}
J.~Fan and D.~S. Williamson, ``From the perspective of perceptual speech quality: The robustness of frequency bands to noise,'' \emph{The Journal of the Acoustical Society of America}, vol. 155, no.~3, pp. 1916--1927, 2024.

\bibitem{rix2001perceptual}
A.~W. Rix, J.~G. Beerends, M.~P. Hollier, and A.~P. Hekstra, ``Perceptual evaluation of speech quality (pesq)-a new method for speech quality assessment of telephone networks and codecs,'' in \emph{2001 IEEE international conference on acoustics, speech, and signal processing. Proceedings (Cat. No. 01CH37221)}, vol.~2.\hskip 1em plus 0.5em minus 0.4em\relax IEEE, 2001, pp. 749--752.

\bibitem{beerends2013perceptual}
J.~G. Beerends, C.~Schmidmer, J.~Berger, M.~Obermann, R.~Ullmann, J.~Pomy, and M.~Keyhl, ``Perceptual objective listening quality assessment (polqa), the third generation itu-t standard for end-to-end speech quality measurement part i—temporal alignment,'' \emph{journal of the audio engineering society}, vol.~61, no.~6, pp. 366--384, 2013.

\bibitem{malfait2006p}
L.~Malfait, J.~Berger, and M.~Kastner, ``P. 563—the itu-t standard for single-ended speech quality assessment,'' \emph{IEEE Transactions on Audio, Speech, and Language Processing}, vol.~14, no.~6, pp. 1924--1934, 2006.

\bibitem{fu2018quality}
S.-W. Fu, Y.~Tsao, H.-T. Hwang, and H.-M. Wang, ``Quality-net: An end-to-end non-intrusive speech quality assessment model based on blstm,'' in \emph{Interspeech}, 2018.

\bibitem{lo2019mosnet}
C.-C. Lo, S.-W. Fu, W.-C. Huang, X.~Wang, J.~Yamagishi, Y.~Tsao, and H.-M. Wang, ``Mosnet: Deep learning based objective assessment for voice conversion,'' in \emph{Proc. Interspeech 2019}, 2019.

\bibitem{leng2021mbnet}
Y.~Leng, X.~Tan, S.~Zhao, F.~Soong, X.-Y. Li, and T.~Qin, ``Mbnet: Mos prediction for synthesized speech with mean-bias network,'' in \emph{ICASSP 2021-2021 IEEE International Conference on Acoustics, Speech and Signal Processing (ICASSP)}.\hskip 1em plus 0.5em minus 0.4em\relax IEEE, 2021, pp. 391--395.

\bibitem{yu2021metricnet}
M.~Yu, C.~Zhang, Y.~Xu, S.-X. Zhang, and D.~Yu, ``Metricnet: Towards improved modeling for non-intrusive speech quality assessment,'' in \emph{Interspeech}, 2021.

\bibitem{mittag2021nisqa}
G.~Mittag, B.~Naderi, A.~Chehadi, and S.~M{\"o}ller, ``Nisqa: A deep cnn-self-attention model for multidimensional speech quality prediction with crowdsourced datasets,'' in \emph{Interspeech}, 2021.

\bibitem{cooper2022generalization}
E.~Cooper, W.-C. Huang, T.~Toda, and J.~Yamagishi, ``Generalization ability of mos prediction networks,'' in \emph{ICASSP 2022-2022 IEEE International Conference on Acoustics, Speech and Signal Processing (ICASSP)}.\hskip 1em plus 0.5em minus 0.4em\relax IEEE, 2022, pp. 8442--8446.

\bibitem{ta2024enhancing}
B.~T. Ta, M.~T. Le, V.~H. Do, and H.~T. Thanh~Binh, ``Enhancing no-reference speech quality assessment with pairwise, triplet ranking losses, and asr pretraining,'' in \emph{Proc. Interspeech 2024}, 2024, pp. 2700--2704.

\bibitem{sultana2025noise}
S.~Sultana and D.~S. Williamson, ``A pre-training framework that encodes noise information for speech quality assessment,'' in \emph{Proceedings of the IEEE International Conference on Acoustics, Speech and Signal Processing (ICASSP)}, Hyderabad, India, 2025.

\bibitem{mcshefferty2015just}
D.~McShefferty, W.~M. Whitmer, and M.~A. Akeroyd, ``The just-noticeable difference in speech-to-noise ratio,'' \emph{Trends in hearing}, vol.~19, p. 2331216515572316, 2015.

\bibitem{barker2015third}
J.~Barker, R.~Marxer, E.~Vincent, and S.~Watanabe, ``The third ‘chime’speech separation and recognition challenge: Dataset, task and baselines,'' in \emph{2015 IEEE Workshop on Automatic Speech Recognition and Understanding (ASRU)}.\hskip 1em plus 0.5em minus 0.4em\relax IEEE, 2015, pp. 504--511.

\bibitem{panayotov2015librispeech}
V.~Panayotov, G.~Chen, D.~Povey, and S.~Khudanpur, ``Librispeech: an asr corpus based on public domain audio books,'' in \emph{2015 IEEE international conference on acoustics, speech and signal processing (ICASSP)}.\hskip 1em plus 0.5em minus 0.4em\relax IEEE, 2015, pp. 5206--5210.

\bibitem{visqol}
A.~Hines, J.~Skoglund, A.~Kokaram, and N.~Harte, ``Visqol: an objective speech quality model,'' \emph{EURASIP Journal on Audio, Speech, and Music Processing}, vol. 2015, 12 2015.

\bibitem{5133799}
T.~Manjunath, ``Limitations of perceptual evaluation of speech quality on voip systems,'' in \emph{2009 IEEE International Symposium on Broadband Multimedia Systems and Broadcasting}, 2009, pp. 1--6.

\bibitem{6638348}
A.~Hines, J.~Skoglund, A.~Kokaram, and N.~Harte, ``Robustness of speech quality metrics to background noise and network degradations: Comparing visqol, pesq and polqa,'' in \emph{2013 IEEE International Conference on Acoustics, Speech and Signal Processing}, 2013, pp. 3697--3701.

\bibitem{asupaq}
P.~Manocha, Z.~Jin, and A.~Finkelstein, ``Audio similarity is unreliable as a proxy for audio quality,'' in \emph{Proc.\ Interspeech}, 09 2022, pp. 3553--3557.

\bibitem{baevski2020wav2vec}
A.~Baevski, Y.~Zhou, A.~Mohamed, and M.~Auli, ``wav2vec 2.0: A framework for self-supervised learning of speech representations,'' \emph{Advances in neural information processing systems}, vol.~33, pp. 12\,449--12\,460, 2020.

\bibitem{hsu2021hubert}
W.-N. Hsu, B.~Bolte, Y.-H.~H. Tsai, K.~Lakhotia, R.~Salakhutdinov, and A.~Mohamed, ``Hubert: Self-supervised speech representation learning by masked prediction of hidden units,'' \emph{IEEE/ACM transactions on audio, speech, and language processing}, vol.~29, pp. 3451--3460, 2021.

\bibitem{10626267}
A.~Ravuri, E.~Cooper, and J.~Yamagishi, ``Uncertainty as a predictor: Leveraging self-supervised learning for zero-shot mos prediction,'' in \emph{2024 IEEE International Conference on Acoustics, Speech, and Signal Processing Workshops (ICASSPW)}, 2024, pp. 580--584.

\bibitem{manocha2020differentiable}
P.~Manocha, A.~Finkelstein, Z.~Jin, N.~J. Bryan, R.~Zhang, and G.~J. Mysore, ``A differentiable perceptual audio metric learned from just noticeable differences,'' in \emph{Interspeech}, 2020.

\bibitem{manocha2021cdpam}
P.~Manocha, Z.~Jin, R.~Zhang, and A.~Finkelstein, ``Cdpam: Contrastive learning for perceptual audio similarity,'' in \emph{ICASSP 2021-2021 IEEE International Conference on Acoustics, Speech and Signal Processing (ICASSP)}.\hskip 1em plus 0.5em minus 0.4em\relax IEEE, 2021, pp. 196--200.

\bibitem{nomad}
A.~Ragano, J.~Skoglund, and A.~Hines, ``Nomad: Unsupervised learning of perceptual embeddings for speech enhancement and non-matching reference audio quality assessment,'' in \emph{ICASSP 2024 - 2024 IEEE International Conference on Acoustics, Speech and Signal Processing (ICASSP)}, 2024, pp. 1011--1015.

\bibitem{ragano2024scoreq}
------, ``{SCOREQ}: Speech quality assessment with contrastive regression,'' in \emph{The Thirty-eighth Annual Conference on Neural Information Processing Systems}, 2024.

\bibitem{serra2021sesqa}
J.~Serr{\`a}, J.~Pons, and S.~Pascual, ``Sesqa: semi-supervised learning for speech quality assessment,'' in \emph{ICASSP 2021-2021 IEEE International Conference on Acoustics, Speech and Signal Processing (ICASSP)}.\hskip 1em plus 0.5em minus 0.4em\relax IEEE, 2021, pp. 381--385.

\bibitem{chen2020simple}
T.~Chen, S.~Kornblith, M.~Norouzi, and G.~Hinton, ``A simple framework for contrastive learning of visual representations,'' in \emph{International conference on machine learning}.\hskip 1em plus 0.5em minus 0.4em\relax PmLR, 2020, pp. 1597--1607.

\bibitem{acodl}
A.~Ragano, E.~Benetos, M.~Chinen, H.~Martinez, C.~Reddy, J.~Skoglund, and A.~Hines, ``A comparison of deep learning mos predictors for speech synthesis quality,'' in \emph{2023 34th Irish Signals and Systems Conference (ISSC)}, 06 2023, pp. 1--6.

\bibitem{mrh}
H.~Martinez, A.~Ragano, and A.~Hines, ``Exploring the influence of fine-tuning data on wav2vec 2.0 model for blind speech quality prediction,'' in \emph{Interspeech}, 09 2022, pp. 4088--4092.

\end{thebibliography}

\end{document}